\documentstyle[aps,epsfig,epsf]{revtex}

\newcommand{\la}{\langle}
\newcommand{\ra}{\rangle}

\newcommand{\be}{\begin{equation}}
\newcommand{\ee}{\end{equation}}
\newcommand{\bea}{\begin{eqnarray}}
\newcommand{\eea}{\end{eqnarray}}

\newcommand{\der}[2]{\frac{\partial #1 }{\partial #2}}
\newcommand{\T}{\vec{T}(t)}
\newcommand{\dT}{\dot{\vec{T}}(t)}   
\begin{document}
\twocolumn[
 \hsize\textwidth\columnwidth\hsize
 \csname@twocolumnfalse\endcsname

\draft
\setlength{\topmargin}{0.6cm}
\title{Noise-assisted classical adiabatic pumping in a
symmetric periodic potential}

\author{O. Usmani$^1$, E. Lutz$^{1,2}$ and M. B\"uttiker$^1$}

\address{$^1$D\'epartement de
Physique Th\'eorique,
Universit\'e de Gen\`eve 24, quai Ernest Ansermet CH-1211 Gen\`eve 4}

\address{$^2$Sloane Physics Laboratory, Yale University, PO. Box
208120, New Haven, CT 06520-8120, USA}

\date{\today}

\maketitle

\begin{abstract}
 We consider a classical overdamped Brownian particle moving in a
 {\it symmetric} periodic potential.
 We show that a net particle flow can be produced by  
 adiabatically changing two external  periodic
 potentials  with a  phase difference $\varphi$ in time and $\chi$ in space. 
 The classical pumped
 current is found to 
 be independent of the friction and to 
 vanish both in the limit of low and high temperature.  
 Below a critical
 temperature,   adiabatic pumping appears  
 to be   
 more efficient than transport due to a
 constant external force.
\end{abstract}


]

\section{Introduction}

Recently there has been a considerable interest in 
small amplitude adiabatic pumping in mesoscopic electrical conductors \cite{swi99}. 
Two oscillating, out-of-phase perturbations are applied 
which lead to small distortions of the shape of the system. 
As a consequence a directed current is generated \cite{ZSA99}.
The pumped current 
is a consequence of quantum interference effects.  
An elegant formulation of the problem has been achieved  
by Brouwer \cite{Brouwer98} based on the modulation of the emissivities 
of the system \cite{BTP94}. 
Inelastic scattering does not suppress the 
pumped current but introduces an additional more classical 
contribution to the pumped current due to rectification \cite{mos01,Cremer}. 
For a broader view of this very active field we refer to a few recent 
works \cite{avron,aleks,mos02,Poli,Wang}. 

Adiabatic pumping is of general interest due to the fact 
that only very slow perturbations are required: furthermore 
if the amplitudes are small the system remains at all times 
close to the stationary equilibrium state. Thus parametric adiabatic pumping 
can be viewed as a tool to investigate the near equilibrium 
properties of a system.  Since perturbations can be applied 
locally such an investigation gives information 
on the system which can not be obtained through the application 
of global and stationary forces.  

It is the purpose of this work to complement 
the quantum mechanical discussions mentioned above 
and to investigate parametric pumping for a purely classical 
system. We consider particles subject to damping and 
thermal noise in a symmetric periodic potential $V_{0}(x) = V_{0}(-x)$. 
In addition to the static potential $V_{0}(x)$ two small 
amplitude time-dependent oscillatory potentials act on the particles. 
The perturbations we consider are periodic in time and are periodic 
in space with the same period as the static potential. 
We investigate the case where 
the perturbations have a {\it double} phase-difference both in {\it time} 
and {\it space}. 
As in the quantum case, a directed current is generated
which is proportional to the frequency of the oscillating 
perturbations and proportional to the product 
of their amplitudes. A directed current results 
for almost all type of perturbations, unless these 
perturbations have a special symmetry (phase-differences equal to a multiple 
of $\pi$). Interestingly, 
for the small amplitude perturbations considered here, 
the thermal noise is essential: the pumped current vanishes 
in the zero-temperature limit, is maximal at some intermediate 
temperature, and vanishes in the high temperature limit. 

A symmetry breaking of the system is necessary to generate 
a directed current. In the noise assisted parametric pumping process
discussed here  
the symmetry is broken not with the help of the static potential 
$V_{0}(x)$ but through the perturbations applied to the system. 
Non-equilibrium state-dependent noise with the same 
period as the potential but out of phase with a symmetric static 
potential also leads to a directed current \cite{but87,vanK,lan88,bla98,seik}. 
Thus this is an 
example of directed motion in a symmetric potential for which the symmetry 
is broken not by the static potential but only by the non-equilibrium noise. 
Similarly directed motion can be obtained in systems with a spatially
symmetric potential but with a friction constant which is state 
dependent \cite{luch,jay}. 
We emphasize the symmetry of the static problem, since typically, 
the recent literature has emphasized directed transport in systems 
in which already the static 
potential \cite{jul97,ast97,rei01} is asymmetric $V_{0}(x) \ne V_{0}(-x)$.
The examples discussed here and in Refs. \onlinecite{but87,vanK,lan88,bla98,seik}
demonstrate that 
static symmetric breaking, i. e.  the consideration 
of a {\it ratchet} potential \cite{mag93,ast94,mil94,doe94,bar95}
is not necessary,  
if either non-equilibrium 
noise, or perturbations applied to the system act in a symmetry breaking way.

Quasi-adiabatic perturbations of particles subject to the 
Smoluchowski equation have recently been investigated by Parrondo \cite{par98}. 
Below we present a discussion of small amplitude parametric 
pumping which closely follows the discussion by Parrondo \cite{par98}.

\section{Parametric Pumping}

The overdamped motion of a classical particle in an external
potential and subjected to thermal noise
 is governed by the Smoluchowski equation for the probability
 density $\rho(x,t)$,
\bea \label{eq1} \der{}{t} \rho(x,t) &=&   \mu \der{}{x} \left[
\der{V(x,\T)}{x} +
 \frac{1}{\beta} \der{}{x}\right]  \rho(x,t)
\nonumber\\
  &=& - \der{}{x}
 J(x,t)\rho(x,t)\ ,
\eea where $\mu$ is the mobility, $\beta$ the inverse temperature
and $J(x,t) = -\mu V'(x,\T) -\mu
 kT\partial/\partial x$ is the current operator. Here the prime denotes derivative with respect to
 $x$. We consider a  total potential $V(x,\T)$ that is written as a sum of a symmetric periodic
 potential
\be
\label{pot} V_0(x) = V_0 [1-\cos(2\pi x/a)] \ee with  period $a$,
plus a  perturbation \bea \Delta V(x,\T ) = X_1(x) T_1(t) + X_2(x)
T_2(t) . \eea Here $X_1(x)$ and $X_2(x)$ are arbitrary spatial
functions with period $a$ and similarly $T_1(t)$ and $T_2(t)$ are
arbitrary functions of time with period $2 \pi/\omega$. We
consider two special examples with purely harmonic driving. In
both examples the time--dependent external perturbation is
composed of two sinusoidal potentials with amplitude $V_A$ and
$V_B$. In the first example the spatial functions act over the
entire period, 
\bea \label{ppot1} \Delta V(x,\T) = &-& V_A
\cos(2\pi x/a)\cos(\omega t)\nonumber\\ &-& V_B \cos(2\pi
x/a+\chi)\cos(\omega t+\varphi) 
\eea 
with a phase difference
$\chi$ in space and a phase difference $\varphi$
in time. In the second example driving is spatially
localized at two arbitrary points $x_1$ and $x_2$ in the interval
$[0,a]$, \bea \label{ppot2} \Delta V(x,\T)= &-&  V_A \cos(\omega
t)\delta(x_1-2\pi x/a) \nonumber\\ &-& V_B \cos(\omega
t+\varphi)\delta(x_2-2\pi x/a) \eea where $\delta$  is the Dirac
delta function. In the following we assume that $\Delta V(x,\T)$
changes slowly  in time and that its amplitude is small compared
to the unperturbed potential $V_A , V_B \ll V_0$.

The quantity of prime interest is the mean particle current,
averaged over one period of space and one period of time, \bea
\label{eq2} \la I\ra &=& \frac{\omega}{2\pi a} \int\limits_0^a dx
\int\limits_0^{\frac{2\pi}{\omega}} dt \, J(x,t)\rho(x,t)
\nonumber\\ &=& -\frac{\mu \omega}{2\pi a}  \int\limits_0^a dx
\int\limits_0^{\frac{2\pi }{\omega}} dt  \,
 V'(x,\T)\rho(x,t) \ .
\eea Due to the periodicity of the potential in time and space,
the second term of $J(x,t)$ does not contribute to the current. We
begin by solving the Smoluchowski equation (\ref{eq1}) in the
limit of small driving frequencies, $\omega\rightarrow 0$. In this
limit, the system  remains close to the adiabatic solution
$\rho_0^-(x,t)$, which  is obtained by setting  in Eq.~(\ref{eq1})
$\partial \rho/\partial t =0$. The latter is given by an
equilibrium Boltzmann distribution \bea \label{eq3}
\rho_0^{\pm}(x,t) = Z^{-1}_{\pm}(t) \,e^{-\beta
V(x,\T)}\nonumber\\
Z_{\pm}(t) = \int\limits_0^a dx \, e^{\pm \beta V(x,\T)}  \ , \eea
and therefore does not yield any current. (Here we have in
addition to the adiabatic solution $\rho_0^{-}(x,t)$ introduced
$\rho_0^{+}(x,t)$ for later reference). Since the adiabatic
solution is not associated with a current flow, we need to find
the correction to this solution to determine the current. We
expect that the correction to the adiabatic solution is of the
order of the variation rate of the perturbation \cite{par98},
which, for the potential  $\Delta V(x,\T)$ we consider here, is
given by $\omega$. We thus seek a solution of the Smoluchowski
equation of the form
\be
\label{eq4} \rho(x,t) \simeq \rho_0^-(x,t) + \dT\, \vec{\nu}(x,\T)
\ . \ee The correction  $\dT\vec{\nu}(x,\T)$, of order $\omega$,
to the adiabatic solution $\rho_0^-(x,t)$ gives rise to  the
non--vanishing particle current. Inserting the ansatz (\ref{eq4})
into the Smoluchowski equation (\ref{eq1}) and neglecting the time
derivative of the correction, which is of the order $\omega^2$, we
arrive at
\be
\label{eq5} \mu \der{}{x}\left[ \der{V(x,\T)}{x} +
\frac{1}{\beta}\der{}{x}\right]\vec{\nu}(x,\T) = \vec{\nabla}_{\T}
\rho_0^-(x,t) \ . \ee This second-order partial differential
equation for $\vec{\nu}(x,\T)$ has to be solved with periodic
boundary conditions, $\vec{\nu}(0,\T)=\vec{\nu}(a,\T)$, and the
condition that the integral of $\vec{\nu}(x,\T)$ along the
interval $[0,a]$ vanishes. This second condition follows from the
normalization of $\rho (x,t)$ over one (spatial) period. We find
\bea \label{eq6} &&\vec{\nu}(x,\T)= \vec{C}_1 \,e^{-\beta V(x,\T)}
\int\limits_0^x dy \, e^{\beta V(y,\T)}\nonumber\\ &+& \frac{\beta
}{\mu
 } e^{-\beta V(x,\T)} \int\limits_0^x dy\, e^{\beta V(y,\T)}\int\limits_0^{y}
 dz\,\vec{\nabla}_{\T} \rho_0^-(z,t) \nonumber\\
 &+& \vec{C}_2 \,e^{-\beta V(x,\T)}\ ,
\eea where $\vec{C}_1(t)$ and $\vec{C}_2(t)$ are two vectors of
integration constants. Explicitly, $\vec{C}_1(t)$ is given by
\be
\vec{C}_1(t)= -\frac{\beta }{\mu } \, \int\limits_0^a dx
\,\rho_0^+(x,t) \int\limits_0^x dy \,
\vec{\nabla}_{\T}\rho_0^-(y,t) \ , \ee where $\rho_0^{\pm}(x,t)$
is given by Eq.~(\ref{eq3}). The solution  (\ref{eq4}) of the
Smoluchowski equation is then obtained by combining
Eqs.~(\ref{eq3}) and (\ref{eq6}). Using the above solution, the
mean current (\ref{eq2}) can be easily calculated, \bea
\label{eq7} \la I \ra &=& -\frac{\mu  \omega^2}{2\pi \beta
\omega_0} \int\limits_0^{\frac{2\pi}{\omega}} dt \,
\dT\vec{C}_1(t)\nonumber\\ &=& \frac{\omega}{2\pi}
\int\limits_{\vec{T}(0)}^{\vec{T}(\frac{2\pi}{\omega})} d\vec{T}
\int\limits_0^a dx \rho_0^+(x,t) \int\limits_0^x dy
\vec{\nabla}_{\T}\rho_0^-(y,t)  . \eea We now take the case $
\Delta V(x,\T)=T_1(t)X_1(x)+T_2(t)X_2(x)$. With
$\vec{\nabla}_{\T}=\partial/{\partial T_1}+\partial/{\partial
T_2}$ and Green's theorem, Eq.~(\ref{eq7}) can be rewritten as :
\bea \label{eq7b} \la I \ra = \frac{\omega}{2\pi}\int\limits_A
dT_1dT_2\int\limits_0^a dx\int\limits_0^x &&dy
\left(\der{\rho_0^+(x,t)}{T_1}\der{\rho_0^-(y,t)}{T_2}\right.
\nonumber\\
-&&\left.\der{\rho_0^+(x,t)}{T_2}\der{\rho_0^-(y,t)}{T_1}\right).
\eea This is the key result of our paper. Eq.~(\ref{eq7b}) gives
the pumped current in terms of the derivative of the adiabatic
solution  $\rho_0^{-}(x,t)$ and its companion $\rho_0^{+}(x,t)$.
It is valid for slowly time--varying periodic potentials $\Delta
V(x,\T )$ of arbitrary shape and arbitrary strength. The
expression  Eq.~(\ref{eq7b}) is most useful for the case of small
amplitude perturbations. In this case the derivatives of
$\rho_0^{\pm}(x,t)$ can be evaluated at zero amplitude and the
integral over the area enclosed by the path $\int dT_1 dT_2$ is
simply a multiplying factor.

\section{Example with global driving}

We now specialize to the small amplitude regime. To do so, we can
set $T_1 = T_2 = 0$ in the integrand of Eq.~(\ref{eq7b}). For the
particular potential $\Delta V(x,\T)$ introduced in
Eq.~(\ref{ppot1}) the area enclosed by the pumping path is
$\int\limits_A dT_1dT_2V_AV_B\sin\varphi \sin\chi $ and a
calculation leads to the pumped current
\be
\label{eq9} \la I \ra=\frac{1}{2 \pi V_0^2}\,\beta V_0
\frac{I_1(\beta V_0)}{I_0^3(\beta V_0)}\, \omega V_AV_B
\sin\varphi \sin\chi \ , \ee where $I_0(x)$ and $I_1(x)$ are the
hyperbolic Bessel functions of order zero and one, respectively.
Equation (\ref{eq9}) exhibits the main features of adiabatic
pumping. We see that the adiabatically pumped current $\la I \ra$
is linear in  the pumping frequency $\omega$ and the amplitudes
$V_A$ and $V_B$ of the two external potentials. The current is
proportional to the sines of the temporal and spatial phase
differences. An important consequence of Eq.~(\ref{eq9}) is that
the current vanishes if either $\varphi$ or $\chi$ is a multiple
of $\pi$. This shows that a {\it double} phase difference, both in
time and in space is necessary in order to rectify the noise. More
generally, it can be shown that if the perturbation is written as
product $\Delta V(x,\T) = X(x)T(t)$ with $X(x)$ periodic in space
and $T(t)$ periodic in time, there is no pumped current, for any
amplitude. An interesting situation, which offers a simple
physical interpretation, is the one for which the current is
maximum. This happens  when  $\varphi,\chi=\pm\frac{\pi}{2}$. By
taking $|V_A|=|V_B|$,  the perturbation can  be rewritten in the
form
\be
\Delta V(x,\T)= V_A \cos (2\pi  x/a \pm \omega t), \ee the sign
being determined by the relation between $\varphi$ and $\chi$ and
between $V_A$ and $V_B$. The maximum current is hence generated by
a traveling wave potential. It is to be expected that a traveling
wave potential is a particularly efficient way of generating a
current \cite{trav1,trav2}.

\begin{figure}[t]
\vspace{3mm} 
\centerline{\epsfxsize=7.3cm \epsffile{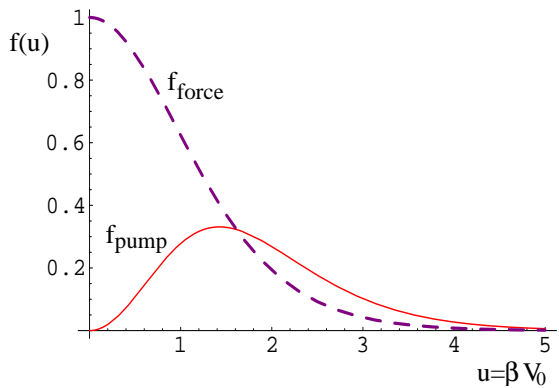}}
\vspace{3mm}
 \nopagebreak
\caption{Temperature dependence $f(u=\beta  V_0)$, of the
adiabatically pumped current
 (\ref{eq9}) (solid line) and the current  (\ref{eq11})
 generated by a constant external force
 $F$ (dashed line).}
\label{fig1}
\end{figure}

Let us now examine the temperature dependence of the particle
current. It is given by the function
\be
\label{eq10} f_{\mbox{\small pump}}(u=\beta V_0
)=u\frac{I_1(u)}{I_0^3( u)} \ . \ee This function has been plotted
in Fig.(\ref{fig1}) (solid line). We observe  that
$f_{\mbox{\small pump}}(u)$ vanishes both in the limit of low and
high temperature and that it reaches a maximum at $u \simeq
1.426..$. This behavior can be understood as follows. At very low
temperature (large $u$), thermal activation is negligible. Since
the perturbation $\Delta V(x,\T)$ is furthermore very small, the
particle remains trapped in the minima of the bare potential
$V_0(x)$ and there is no transport. For the small amplitude
perturbations considered here this clearly demonstrates that there
is no classical pumping without thermal noise. In the opposite
limit of high temperature (small $u$), thermal fluctuations
dominate and we have simple (symmetric) Brownian diffusion with
zero average displacement. The maximum pumped current corresponds
to a thermal energy of the order of the potential energy.

It is further instructive to compare the adiabatic pumped current
(\ref{eq9}) with the current created  by a small constant (time
and space independent) external force $F$. In this case the
overdamped Brownian particle experiences the potential, $ V(x)
=V_0(x) -Fx$. A similar calculation of the current, up to first
order in $F$, yields the following expression  (see also
\cite{ris89}),
\be
\label{eq11} \la I \ra=\frac{\mu F}{aI_0^2(\beta V_0)} \ . \ee The
temperature dependence of this force induced current, given by
$f_{\mbox{\small force}}(u)=I_0^{-2}(u)$, is shown in
Fig.(\ref{fig1}) (dashed line). Here, in contrast to adiabatic
pumping, the current is maximum at very high temperature. This is
due to the presence of a small but finite slope in the potential.
In the low temperature limit both currents are suppressed.
\noindent We  notice, however, that for the force induced current,
$f_{\mbox{\small force}}(u) \sim u\, e^{-2u}$, the decay with
temperature is asymptotically  {\it faster} than for  the pumped
current, $f_{\mbox{\small pump}}(u) \sim u^2\, e^{-2u}$. As a
consequence, there is a critical temperature $\beta_c$ below which
the pumped current is larger than the current generated by $F$. In
this regime, adiabatic  pumping is a more efficient transport
mechanism than applying an external force. The value of the
critical temperature $\beta_c$ depends on both the area enclosed
by the pumping path, the frequency and $F$.

\section{Example with localized driving}

\begin{figure}[t]
\vspace{3mm} 
\centerline{\epsfxsize=7.3cm \epsfbox{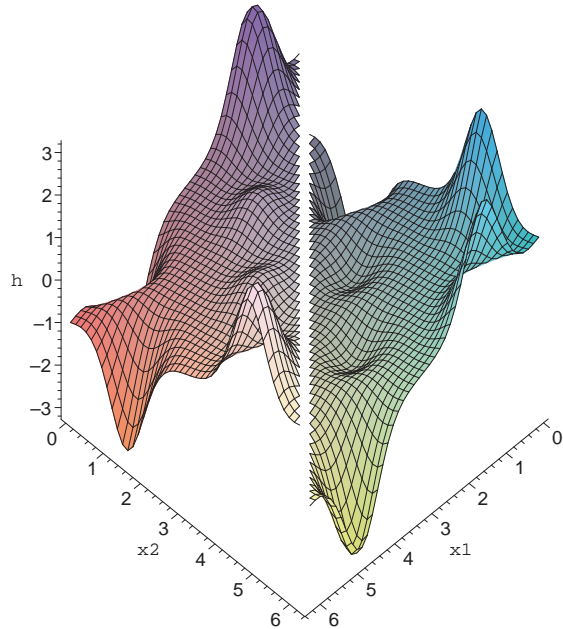}} 
\vspace{3mm}
\caption{ $h(1,x_1,x_2)$, which describes the variation of the
pumped current with respect to the positions $x_1$ and $x_2$ of
the delta perturbations. There is no pumped current for $x_1 =
x_2$. } \label{fig2}
\end{figure}

In the example discussed above the perturbation potential is
extended through the entire spatial period of the system. Is this
a necessary condition, or is a spatially localized perturbation
sufficient to generate a pumping current? To answer this question
we consider a perturbation given by Eq. (\ref{ppot2}). As for the
previous example, we can compute the current from Eq.~(\ref{eq7b})
to obtain:
\be
\la I\ra=-\frac{\omega V_AV_B\sin\varphi}{8\pi^3V_0^2}
\left(\frac{\beta V_0}{I_0\left(\beta V_0\right)}\right)^2 h(\beta
V_0,x_1,x_2), \ee where $h(\beta V_0,x_1,x_2)=f(\beta
V_0,x_1,x_2)+f(-\beta V_0,x_1,x_2)$, and \bea f(\beta
V_0,x_1,x_2)= &&e^{(2\beta
V_0\sin(x_1+x_2)\sin(x_1-x_2))}\nonumber\\
&\times&\left(\theta(x_1-x_2)+\int\limits_{x_1}^{x_2-\pi} dx
\tilde{\rho}_0^-(x)\right), \eea with
$\tilde{\rho}_0^-(x)={\rho}_0^-(x,T_1=0,T_2=0)$. $\theta$ is the
Heaviside step function. If we exchange $x_1$ and $x_2$, we change
the sign of the current, since it corresponds to a change of the
sign of the temporal phase difference. The temperature dependence
is similar to the one encountered previously (i.e. the current
vanishes for both zero and infinite $\beta V_0$). The interesting
dependence of the current on $x_1$ and $x_2$ is determined by
$h(\beta V_0=1,x_1,x_2)$ which is plotted in Fig.(\ref{fig2}). No
current is generated if $x_1 = x_2$. The pumped current (the
function $h(\beta V_0,x_1,x_2)$) is discontinuous along the line
$x_1 = x_2$. Of course this discontinuity disappears if instead of
a delta functions slightly broadened functions are used to
describe the perturbation.

The reason that the localized perturbation leads to a pumped
current is due to the normalization of the distribution function.
As a consequence even a localized perturbation generates a
non-local response. 

Fig. (\ref{fig2}) gives an overview of the pumping currents that
can be achieved with the two localized perturbations. To gain
further insight it is useful to consider several cuts through Fig.
(\ref{fig2}).

\begin{figure}[t]
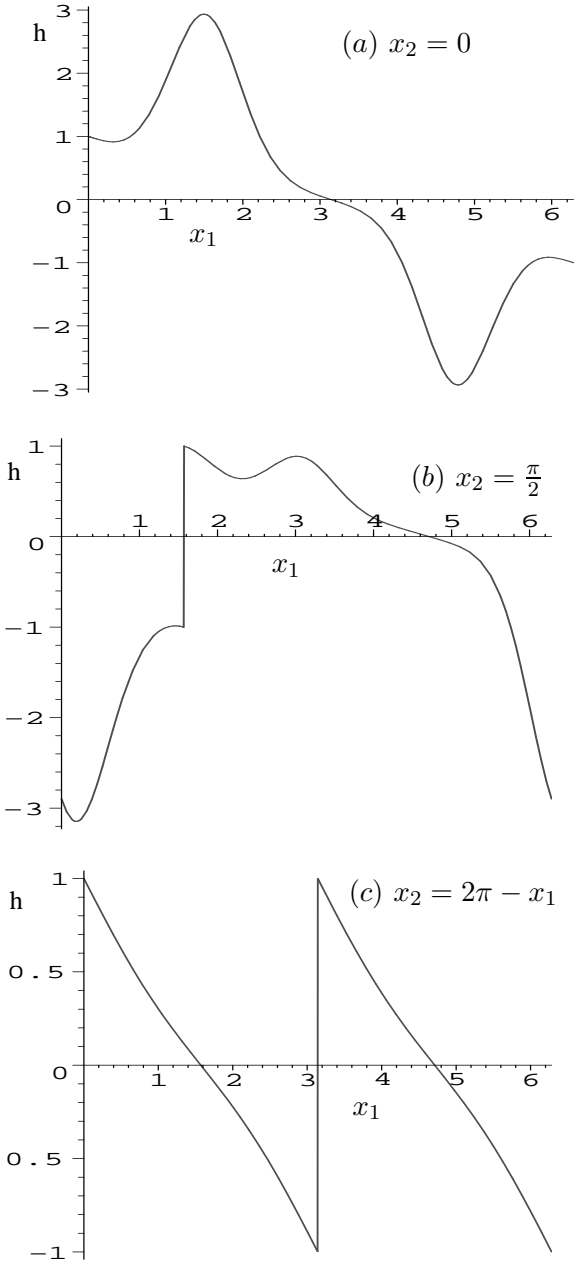

\input{fig3a.pstex_t} \vspace{.5cm}\\
\input{fig3b.pstex_t} \vspace{.5cm}\\
\input{fig3c.pstex_t} \vspace{3mm}
\caption{Variation of the pumped current of the localized
parametric pumping model. Shown is the function $h$ for (a) $x_2 =
0$, (b) $x_2 = \frac{\pi}{2}$ and (c) $x_2 = 2 \pi - x_1$. }
\label{fig3}
\end{figure}

In Fig. (\ref{fig3}) (a) we keep the position of one perturbation
at a fixed location $x_2 = 0$ (at the potential minimum) and
consider the pumping current as a function of the position $x_1$
of the other perturbation. The pumped current is then maximal if
$x_1$ is at the inflection point of the potential ($x_1 = \pi/2$).
There is no current if the perturbation is located at the maximum
of the potential. As $x_1$ increases further the current direction
is reversed. As $x_1$ passes through zero the current jumps from a
negative value to a positive value. Note that in this case the
pumped current is anti-symmetric around $x_1 = \pi$ since one
perturbation is located at a symmetry point ($x_2 = 0$) of the
potential.

In Fig. (\ref{fig3}) (b) we keep one perturbation at the
inflection point of the potential $x_2 = \pi/2$. As a function of
$x_1$ the pumped current jumps at  $x_1 = \pi/2$. The current
decreases as $x_1$ increases past $x_2 = \pi/2$ goes through a
local minimum, reaches a local maximum for $x_1$ just before $x_1
= \pi$. The maximal pumped current is achieved if $x_1$ is a
little to the right of the local potential minimum. Note that in
this case we have no symmetry around $\pi/2$ and neither are the
two directions of current equivalent.

In Fig.  (\ref{fig3}) (c) we consider the case where the two
perturbations are symmetrically located around the potential
maximum $\pi$. The current then jumps at $x_1 = x_ 2 = \pi$ and is
antisymmetric as a function of $x_1$ around this point.

For global driving we found that the current is a sinusoidal
function of the spatial phase difference between the two
perturbations. In contrast, in the local driving model considered
here the pumped current depends not in a simple manner only on the
spatial distance between the two perturbations but also on their
absolute locations in the interval. This leads to the much more
complicated behavior depicted Fig. (\ref{fig2}) and Fig.
(\ref{fig3}) (b).

\section{Discussion}

We first point to the remarkable fact that the pumped current 
is independent of the mobility $\mu$. 
This should be contrasted with the linear dependence 
on $\mu$ of a current generated with the 
help of a constant force. That the pumped current 
is indeed independent of $\mu$ can easily be seen 
by considering Eq. (\ref{eq5}). The solution of this equation determines
the correction  $\nu$ of the density to the adiabatic solution: 
This correction is proportional to $\mu^{-1}$. Since the current operator 
is proportional to $\mu$ the resulting net current is independent 
of $\mu$. We emphasize that the pumped current is independent 
of $\mu$ not only for the particular models considered here but 
quite generally for all adiabatic perturbations. 
 
In summary, we have shown that a directed current can be generated
in a symmetric periodic potential by adiabatically modulating two
small external potential parameters. We have investigated a model
with global (spatially periodic) perturbations and a model with
localized perturbations. For the model with global perturbations
we find that a double --- temporal and spatial
 ---  phase difference
is necessary to generate a current. The maximum pumped current is
obtained for a temperature $kT$ of the order of the potential
height $V_0$ and for  a perturbative potential corresponding to  a
traveling wave ($\varphi,\chi =\pm \pi/2$). Similarly for the
model with localized perturbations we find a pumped current unless
the two perturbations are located at the same position or have a
spatial difference of $\pi$. Pumping arises through the subtle
interplay between  thermal fluctuations and  cyclic variations of
the potential. It therefore disappears in the limit of low and
high temperature when either the potential or the thermal energy
becomes predominant.  We have demonstrated the existence of a
potential dependent critical temperature below which adiabatic
pumping is a more efficient than applying a
small constant external force. The work presented here can be
extended in different directions: Systems with open boundary
conditions, several space dimensions and inertial effects are
possible subjects for further research.

We thank M. Moskalets and A. Alekseev for discussions. 
This work was supported by the RTD network 
Nanoscale Dynamics, Coherence and Computation and by 
the Swiss National Science Foundation.

\end{document}